# Range Extension in IEEE 802.11ah Systems Through Relaying

Enis Kocan[1] · Bojan Domazetovic[2] · Milica Pejanovic-Djurisic[1]

**Abstract** – A great number of Internet of Things (IoT) and machine-to-machine (M2M) based applications, which are telecommunication areas with the highest foreseen growth in the future years, require energy efficient, long range and low data rate wireless communication links. In order to offer a competitive solution in these areas, IEEE 802.11 standardization group has defined the "ah" amendment, the first sub-1GHz WLAN standard, with flexible channel bandwidths, starting from 1MHz, up to 16MHz, and many other physical and link layer improvements, enabling long-range and energy efficient communications. However, for some regions, like Europe, the maximum transmitted power in dedicated frequency band is limited to only 10mW, thus disabling the achievement of ranges which would be close to targeted of up to 1km. In this paper we examine possibilities for range extension through implementation of half-duplex decode-and forward (DF) relay station (RS) in communication between an access point (AP) and an end-station (ST). Assuming a Rician fading channel between AP and RS, and a Rayleigh fading channel on the RS – ST link, we analytically derive results on achievable ranges for the most robust modulation and coding schemes (MCSs), both on downlink (DL) and uplink (UL). Analyses are performed for two different standard adopted deployment scenarios on the RS – ST link, and variable end-to-end link outage probabilities. Moreover, we have analyzed whether the considered most robust MCSs, known for supporting the longest range, but the lowest data rates, can meet the defined requirement of at least 100kb/s for the greatest attainable AP - RS - ST distances. We examine data rate enhancements, brought by coding and using of short packets, for both DL and UL. Finally, we present bit error rate (BER) results, obtained through simulations of a dual-hop DF IEEE 802.11ah relay system for the considered MCs. All presented results confirm that IEEE 802.11ah systems, through deployment of relay stations, become an interesting solution for M2M and IoT based applications, due to flexibility they offer in many aspects, meeting requirements for wide transmission ranges in such applications.

Enis Kocan, e-mail: enisk@ac.me
B. Domazetovic, e-mail: bojan.domazetovic@rdc.co.me
M. Pejanovic-Djurisic, e-mail: milica@ac.me
[1] University of Montenegro, Faculty of Electrical Engineering, Podgorica
[2] Broadcasting center Montenegro, Podgorica, Montenegro

**1 Introduction**

For many decades, development of different telecommunication systems has been leading towards the goal of achieving man-to man communications anywhere, anytime and with anybody. During the last decade of the 20$^{th}$ century, with expansion of Internet, this paradigm has slightly shifted towards achieving man-to-machine communications. Nowadays, our generations are witnessing the next conceptual change in telecommunications, with the focus on solutions for direct machine-to-machine (M2M) communications, as well as on provision of Internet connection for exponentially increasing number of different "smart" devices (IoT – Internet of Things). Analyses forecast that, by 2020, tens of billions of different devices will be connected, as a part of Internet of Things (IoT) network, [1]. This novel telecommunication concept has significantly affected global telecommunication market, forcing industry and service providers to seek for cost-effective and efficient communication solutions for IoT and M2M based applications. For a great number of such applications, the optimal communication solution would be to provide low power, wide range wireless transfer of data. In order to become competitive in this new telecommunication market, WiFi alliance started the work on providing an extended range wireless local area network (WLAN) solution. Thus, in 2010, they organized the work through the IEEE 802.11ah standardization group, with the goal to design the first sub-GHz WLAN standard, [2].

In great number of M2M and IoT based applications, end-nodes will generate sporadic, low data rate traffic, but there will be hundreds, or thousands of nodes connected to single gateway/access point/base station, depending on the communication system used. Examples of such applications are data acquisition from different type of sensors, performing environmental/plant/animal/traffic monitoring, smart metering, controlling different actuators, etc. Thus, the IEEE 802.11ah working group aimed at specifying a network that will enable connection of up to 6000 devices to a single access point (AP), with communication range of up to 1km, and data rates of at least 100kb/s, [3].

Following on the existing standard and new requirements, IEEE 802.11 has developed an amendment, IEEE 802.11ah, defined as a sub-1GHz license-exempt OFDM (orthogonal frequency division multiplexing) based standard for future M2M communications. The major improvements, introduced on physical and link layers, have enabled a flexible WLAN standard, which supports a variety of application scenarios, starting from common WLAN applications, up to providing a long range communication links for a large number of energy constraint devices. Unlike the other IEEE 802.11 standards (a, b, g, n, ac) which operate in 2.4GHz and 5GHz frequency bands and are optimized for a limited number of high data rate devices, 802.11ah supports network with a large number of low data rate devices enabling much longer transmission range, by using: propagation benefits of operation in sub-1-GHz spectrum, flexible channel bandwidths (from 1MHz up to 16MHz), robust modulation and coding schemes (MCS), short medium access control (MAC) frame, increased sleep time, target wakeup time, etc., [2]. Thus for example, link budget gain of 24.5dB can be attained in the IEEE 802.11ah system, operating over 1MHz channel bandwidth at 900MHz band, when comparing with the IEEE 802.11n system, operating at 2.4GHz, [2].

Even before the first draft of IEEE 802.11ah standard appeared, which happened in October 2013, research community had expressed interest in performance of the upcoming WLAN standard, and this interest is now moving towards novel solutions for its performance improvements [4] – [12]. Validity of the proposed IEEE 802.11ah propagation path loss models for urban smart grid applications is tested in [4]. The achievable data rates in the case of shadow fading presence in outdoor environments, for different MCSs, packet lengths and link outage probabilities, are examined in [5]. Channel occupancy, packet delivery ratio, delay, energy consumption and data rates for two indoor (smart metering and industrial automation) and two outdoor (agriculture monitoring and animal monitoring) application scenarios are analysed in [6]. Throughput of the IEEE 802.11ah system, with medium access control (MAC layer) mechanisms included, is presented in [7]. Comprehensive overview of the challenges and general findings in achieving long-range WLAN communications is given in survey [8]. In case where only path loss effects are taken into account, achievable transmission ranges for various MCS schemes and transmission powers, as well as throughput performance for different MAC and MCSs, are evaluated in [9], showing that maximum range is up to 1550m for MCS10 and 1W transmission power. Implementing simulation model for the scenario with Rayleigh multipath fading statistics, previously we have evaluated the achievable ranges on downlink (DL) and BER performances of the most robust MCSs, for different transmission powers, [10]. The obtained results have confirmed that even in a Rayleigh multipath fading environment, the targeted range of 1km can be achieved for transmission power of 1W. However, depending on the regions and country regulations, maximum transmission power for IEEE 802.11ah systems may be limited to only 10mW, like it is in Europe. Results given in [9] and [10] show that such limitation of transmit power prevents achievement of communication ranges close to 1km, even if the most robust MCSs are implemented. Thus, some other techniques should be applied.

In that context, one of the most promising solution for range extension in IEEE 802.11ah systems is implementation of relay stations (RS), [7], which can also bring improved reliability of data transmission in non-line of sight (NLoS) scenarios, and can reduce energy consumption of end-stations (ST). This possibility of employing a dual-hop relay system, where a half-duplex RS applies decode-and-forward (DF) relaying, has been already included in the specification of the IEEE 802.11ah standard, [3], [12]. This means that RS in the first time slot receives information, then performs full decoding, and encodes it again before transmission in the second time slot. RS encompasses two logical entities, access point and end-station, assuming that RS would appear as an access point (AP) to ST, while in the communication with AP, it would behave as ST. There are very few research works in literature dealing with performances of such IEEE 802.11ah relay systems. In [11] the author proposes an algorithm for minimizing sensor energy consumption in IEEE 80211ah relay based network. In [12] authors examine throughput enhancement that a full-duplex inband RS can bring compared to half-duplex RS. However, to the best of authors' knowledge, there are no research results published yet about the benefits that RS implementation can bring to the IEEE 802.11ah system, in terms of range extension and data rate.

Thus, in this paper we examine the achievable transmission ranges for both downlink (DL) and uplink (UL) communications in the IEEE 802.11ah dual-hop relay system, for the most robust MCS schemes, assuming Rayleigh fading statistics on RS-ST link, and Rician fading on AP-RS link. We focus our attention on systems that could be deployed in Europe, which will have a limit on 10mW for the transmit power. Further on, we analyse influence of the expected end-to-end link outage probability ($P_{out\text{-}tot}$) on the achievable range, assuming different deployment scenarios.

We also derive achievable data rates for the assumed fading scenarios in dual-hop DF relay systems, with the goal to examine if the considered MCS schemes, knowing for attaining the longest range, but the lowest data rates, can meet the required data rate of at least 100kb/s. Then, it is analysed how changes in packet sizes, in $P_{out\text{-}tot}$, and implementation of coding, affect these achievable data rates on DL and UL. At the end, using the simulation model developed for the purpose of this research, bit error rate (BER) performances of the IEEE 802.11ah relay system on DL and UL are obtained.

The performance assessment of IEEE 802.11ah DF relay systems presented in this paper is performed with the goal to determine whether the novel WLAN standard can be considered as competitive communication technology for IoT and M2M based applications in Europe. To that aim, the paper is organized as follows. Basic characteristics of physical layer of IEEE 802.11ah systems are presented in Section 2. The achievable ranges in the novel WLAN standard for direct link communication, as well as for the case of RS implementation, are given Section 3. Data rate performances of the IEEE 802.11ah system, both in cases of direct link communication and for dual-hop DF relay communication are derived and analyzed in Section 4. BER performance of the IEEE 802.11ah relay system, obtained through simulations, are presented in Section 5, while concluding remarks are given in Section 6.

## 2 Basic physical layer characteristics

IEEE 802.11ah is based on the OFDM-based technology, which inherits the basic physical layer design from 802.11ac/n standards. The goal of IEEE 802.11ah standard is to offer a robust and efficient solution for the needs of M2M and IoT based applications, requiring in most cases long range and low power wireless communications. For achieving longer communication ranges, besides using sub-1GHz spectrum, new WLAN standard implements narrower channel bandwidth, thus increasing signal to noise ratio (SNR) at the receiver. Compared with the previous IEEE 802.11 standards, where the narrowest used channel bandwidth is equal to 20MHz, the IEEE 802.11ah supports 1MHz, 2MHz, 4MHz, 8MHz and 16MHz channel bandwidths. It is mandatory that all IEEE 802.11ah equipment support channel bandwidths of 1MHz and 2MHz, while the remaining ones are left as options. The number of used sub-carriers for 1MHz channel is 26 per an OFDM symbol (2 pilot tones and 24 data sub-carriers) and for higher channel bandwidths the number of data and pilot tones (fixed, traveled) increases (484 tones for 16MHz bandwidth). The tone spacing between adjacent subcarriers is 31.25kHz for all bandwidth modes, [9]. This makes the inverse/discrete Fourier transform (IDFT/DFT) period equal to 32μs, which is 10 times longer than in 802.11ac systems. The OFDM symbol period is 40μs, comprising 8μs guard interval (GI) (36μs with short GI).

Besides flexibility in channel bandwidths, IEEE 802.11ah systems can choose between 10 different modulation and coding schemes (MCSs), and support up to four spatial streams, which all jointly provides wide variety of possible ranges and data rates they can offer for various applications. Table 1 presents the specified MCSs, with achievable data rates for single spatial stream, depending on the bandwidth used, [7]. It should be noted that, compared to previous IEEE 802.11 standards, MCS10 is the newly introduced MCS, which implies 2x repetition coding, binary phase shift keying (BPSK) modulation and 1/2 coding rate, thus providing the greatest achievable range in 802.11ah systems.

Table 1. MCS and achievable data rates for 1MHz, 2MHz and 16MHz bandwidth, and 1 spatial stream

|  | Modul. | Code Rate | 1MHz (Mb/s) | 2MHz (Mb/s) | 16MHz (Mb/s) |
|---|---|---|---|---|---|
| MCS0 | BPSK | 1/2 | 0.30 | 0.65 | 6.5 |
| MCS1 | QPSK | 1/2 | 0.60 | 1.3 | 13 |
| MCS2 | QPSK | 3/4 | 0.90 | 1.95 | 19.5 |
| MCS3 | 16QAM | 1/2 | 1.2 | 2.6 | 26 |
| MCS4 | 16QAM | 3/4 | 1.8 | 3.9 | 39 |
| MCS5 | 64QAM | 2/3 | 2.4 | 5.2 | 52 |
| MCS6 | 64QAM | 3/4 | 2.7 | 5.85 | 58.5 |
| MCS7 | 64QAM | 5/6 | 3 | 6.5 | 65 |
| MCS8 | 256QAM | 3/4 | 3.6 | 7.8 | 78 |
| MCS9 | 256QAM | 5/6 | 4 | N/A for 1 spat. stream | 86.67 |
| *MCS10 | BPSK | 1/2 | 0.15 |  |  |

*includes 2x repetition mode to increase the range

Minimum receiver sensitivity, or minimum detectable signal (MDS) in 802.11ah systems is going from -98dBm for the binary phase shift keying (BPSK) with 1/2 code rate and repetition coding at 1MHz channel bandwidth (i.e. for MCS10), up to -58dBm for the 256 quadrature amplitude modulation (256QAM) with code rate of 5/6 and at channel bandwidth of 16MHz, [13]. Following MCS10, the next most robust MCS scheme is MCS0, with MDS being equal to -95dBm at the 1MHz channel bandwidth, and -92dBm at the 2MHz channel bandwidth.

The described modifications on physical layer, as well as novel solutions on link layer, will enable fulfilling of technical requirements associated with IEEE 802.11ah adopted use cases, which are namely: sensors and meters, backhaul aggregation, and extended range WLAN and cellular off-loading, [7].

Implementation of IEEE 802.11ah systems for sensing and metering applications represents the most perspective application scenario, comprising smart meters, environmental and agricultural monitoring, smart grids, automation of industrial processes, indoor healthcare/fitness systems, etc. The novel robust MCS and favorable propagation characteristics at sub-1GHz bands, allow creation of a communication approach for wireless sensor networks (WSN), which outperforms other WSN communication solutions operating in licence free bands, like ZigBee and Bluetooth, in terms of throughput and coverage, while remaining very energy efficient (see Table 2), [6].

Table 2. Comparison of 802.11ah with other solutions for WSN

|  | *Zigbee* | *Bluetooth* | *IEEE 802.llah* |
|---|---|---|---|
| *Standard* | IEEE 802.15.4 | IEEE 802.15.1 | IEEE 802.11ah |
| *Frequency band* | EU: 868 MHz<br>North America: 915 MHz<br>Global: 2.4 GHz | 2.4 GHz | Sub-1GHz |
| *Data rate* | 868 MHz band: 20 kb/s<br>915 MHz band: 40 kb/s<br>2.4 GHz band: 250 kb/s | 1 Mb/s | 150kb/s - 347Mb/s |
| *Typical range* | 2.4 GHz band: 10-100 m. | 10-30 m. | 100-1000 m. |
| *TX power* | 1-100 mW | 1-10 mW | 10 mW<$P_{tx}$<1 W (depending on the country's regulations) |

Since IEEE 802.11ah operates in the license-exempt frequency band, the specified operation parameters (frequency bands, maximum effective radiated power-ERP and channel bandwidths-*B*) are different from region to region, or even from country to country, depending on country's regulations (Table 3), [8].

Table 3. IEEE 802.11ah Spectrum, ERP and channel bandwidths

| Geographic area | Frequency [MHz] | ERP [mW] | *B* [MHz] |
|---|---|---|---|
| China | 614-787; 779-787 | 5; 10 | 1 |
| Europe | 863-868.6 | 10 | 1; 2 |
| Japan | 915.9-929.7 | 1; 20; 250 | 1 |
| Singapore | 866-869; 920-925 | 500 | 1; 2; 4 |
| South Korea | 917-923.5 | 3; 10 | 1; 2; 4 |
| United States | 902-928 | 1000 | 1-16 |

As it has been already mentioned, it can be seen from Table 3, that in Europe, China and South Korea, ERP is limited to only 10mW. Thus, it is necessary to examine the achievable ranges of IEEE 802.11ah system in different realistic scenarios, in order to obtain insight in its competitiveness in the mentioned regions.

## 3 Achievable ranges

When analyzing the IEEE 802.11ah standard, the focus is on one of its main features: achievement of long-range communications, targeting distances of up to 1km. To that aim, the standard has incorporated many different physical and link layer solutions, not existing in the other IEEE 802.11 group of standards.

The achievable range of IEEE 802.11ah system can be assessed using the link-budget expressed with:

$$P_{rx} = P_{tx} + G_{tx} - PL(d) + G_{rx} \tag{1}$$

where $P_{rx}$ and $P_{tx}$ represent received and transmitted power, respectively, expressed in dBm. $G_{tx}$ and $G_{rx}$ are transmit and receive antenna gains, respectively, given in dBi, while $PL(d)$ denotes path loss in dB at distance *d*. In this model, we have omitted possible system losses at the transmitter and the receiver. In the channel model adopted for the standard, two possible IEEE 802.11ah outdoor path loss models are used. The first one, denoted as *macro deployment*, assumes that the access point antenna is placed in a position which is 15m above rooftop, and path loss in dB is given by,[4]:

$$PL(d) = 8 + 37.6 \cdot \log_{10}(d), \qquad (2)$$

where $d$ is in meters, and the carrier frequency is 900MHz. Another model is the pico/hot zone deployment outdoor path loss model which assumes that the antenna height is at a roof top level, and path loss is obtained as, [4]:

$$PL(d) = 23.3 + 36.7 \cdot \log_{10}(d) \qquad (3)$$

Fig. 1 presents the level of received power in dBm, for downlink (DL) and uplink communications, for different AP – ST distances. Results representing the received power levels in macro and pico deployment scenarios, for two levels of the AP transmit powers, are given. It is taken that access point (AP) antenna has a gain of 3dBi, while IEEE 802.11ah station (ST) has 0dBi antenna gain, and 0dBm transmit power, when uplink (UL) transmission is considered. In order to assess the achievable ranges, minimum detectable signal (MDS) levels are also given for MCS0, in case where the channel bandwidth is B=2MHz (MDS = -92dBm), then in case where MCS0 is implemented over B=1MHz channel bandwidth (MDS = -95dBm), and for MCS10 (MDS = -98dBm).

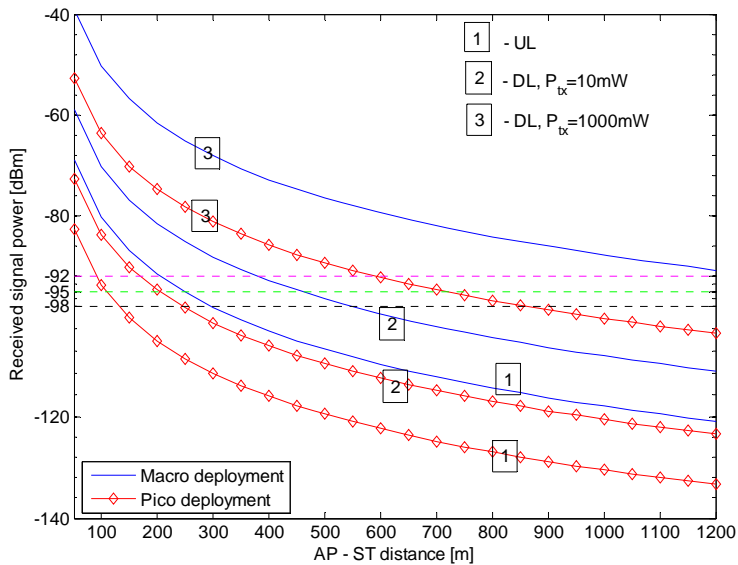

Fig. 1. Achievable ranges, taking into account path loss only

From Fig. 1, it can be seen that the requested range, up to 1km, will be achieved on DL for the macro deployment, in cases where the AP transmit power is equal to 1W, for all three considered MCSs. However, in Europe, where the maximum allowed transmit power of AP is $P_{tx}$=10mW, it is evident that even in the macro deployment scenario the achievable DL range is far below the desired 1km, and it is equal to 550m for MCS10.

For the pico deployment scenario, the greatest achievable range on DL is approximately 850m, and it is achieved for the case of $P_{tx}$=1W and MCS10. For the same deployment scenario and MCS scheme implemented, if $P_{tx}$=10mW, the achievable DL range is approximately 250m. Comparing the results for the macro and pico deployment scenarios, we can see that significantly greater ranges are achieved in the macro deployment scenario, due to smaller path-loss value in this model. This suggests that the AP position will have great influence on the communication range, and it should be placed above rooftops wherever it is feasible.

As it could be expected, the achievable ranges are significantly lower in UL than in any considered DL communication, since the ST transmit power is equal to 1mW. Thus, in the macro deployment scenario, the maximum UL achievable range (for MCS10) is 300m, while in the pico deployment scenario, it is about 140m.

Our previous analysis did not include the influence of fading effect, which additionally reduces the achievable range. Namely, if we take into account the presence of multipath fading on communication link between AP and ST, then another power loss component has to be included in the relation (1), represented with the fade margin (*FM*) term. *FM* has to be introduced in the fading channel analysis, in order to assure that, with a certain probability level, the received signal power will be higher than the minimum detectable signal:

$$P_{rx} = P_{tx} + G_{tx} - PL(d) + G_{rx} - FM , \quad (4)$$

with *FM* also expressed in dB.

Let us assume that the channel experiences Rayleigh fading. Then, the probability density function (PDF) of the signal envelope *R*, is given with:

$$f_R(r) = \frac{r}{\sigma^2} \exp\left(-\frac{r^2}{2\sigma^2}\right); \; r \geq 0 \quad (5)$$

where $E[R^2] = \overline{r^2} = 2\sigma^2$ represents the mean signal power level. Cumulative distribution function (CDF) of the signal envelope is equal to a probability that the signal envelope falls below a certain level, which is equivalent to the link outage probability, $P_{out}$:

$$P(R \leq r) = P_{out} = 1 - \exp\left(-\frac{r^2}{2\sigma^2}\right); \; r \geq 0 \quad (6)$$

In order to determine *FM* for the case of Rayleigh fading channel, it is assumed that the mean signal power is $\overline{r^2} = 2\sigma^2 = 1$. Then, the minimum allowed level of the received signal envelope, $r_{min}$, for the required link outage probability, $P_{out}$, is derived from (6) as:

$$r_{min} = \sqrt{-\ln(1 - P_{out})} \quad (7)$$

Having that the square of $r_{min}$ represents the minimum level of the received signal power, and knowing that the mean signal power is equal to 0dB, *FM* which guarantees the required value of link outage, is obtained as:

$$FM[dB] = 0 - 10 \cdot \log_{10}\left(-\ln(1 - P_{out})\right) \quad (8)$$

Table 4 gives *FM* values for different link outage probabilities, for the Rayleigh fading statistics.

Table 4. *FM* in the Rayleigh fading channel

| $P_{out}$ | *FM* [dB] |
|---|---|
| 1% | 29.99 |
| 5% | 12.89 |
| 10% | 9.77 |
| 20% | 6.51 |
| 40% | 2.92 |

Figure 2 shows achievable ranges for UL and DL communications in the IEEE 802.11ah system, when the link outage probability, due to the presence of Rayleigh multipath fading, is assumed to be 10%. With the additional power loss of 9.77dB in the link budget, which is actually the value of *FM* for $P_{out}$=0.1, it can be seen that achievable ranges are significantly lower than the ones presented in Fig. 1. Thus for example, it is obtained that only in the macro-deployment scenario, for the case of $P_{tx}$=1W in the DL communication and the MCS10 scheme implemented, the communication link can be established between AP and ST which are separated 1km. In the UL communication, maximum achievable ranges are about 80m and 170m, for the pico and macro deployment, respectively.

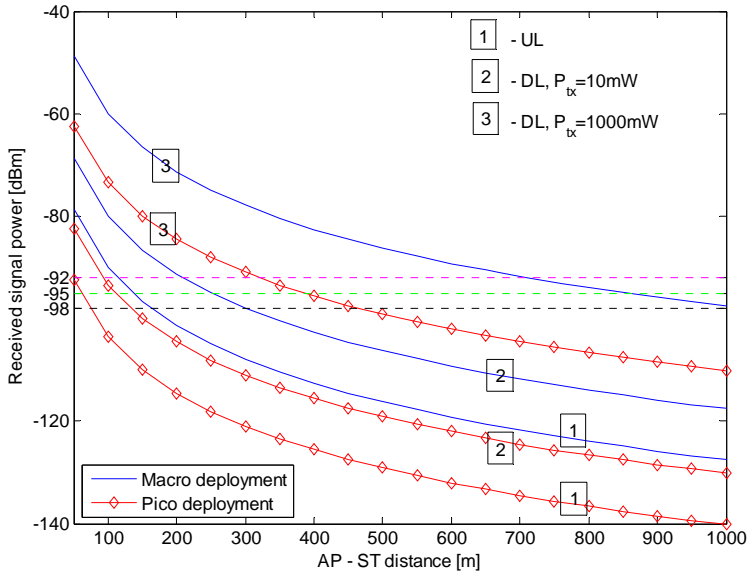

Fig. 2. Achievable ranges for the Rayleigh fading channel ($P_{out}$=0.1)

The results presented so far show that UL communication range is always far below the desired target of 1km, while on DL, only when $P_{tx}$=1W the most robust MCS can ensure, or be very close, this target range. One of the solutions proposed for further range extension in IEEE 802.11ah systems, is implementation of relay stations (RSs). In this analysis, we consider a scenario where all communication processes between AP and ST are performed through RS, meaning that there is no communication link between AP and ST. In a dual-hop relay based communication, where one RS takes part in communication between AP and ST, the end-to-end communication will be in outage if either of links (or hops: AP - RS and RS - S T) is in outage. Thus, due to the cascade connection of these two links, if it is required that the end-to-end communication between AP and ST cannot be in outage for more than $P_{out}$, then each of the two hops must maintain outage performance on $P_{out}/2$.

In our further analysis, focus is on performances of IEEE 802.11ah with implemented DF relays. Moreover, we consider the most critical case for achieving a long-range communication, i.e. the situation when both AP and RS transmit with $P_{tx}$=10mW, while in UL communications, the ST transmit power is equal 1mW. In all relay communication analyses presented in this paper, we assume a realistic scenario where AP is placed in such a position, that the AP - RS link experiences Rician fading, with the Rician *K* factor of 9dB, and the path-loss on this link is always calculated assuming macro deployment scenario. RS - ST link has Rayleigh fading statistics, and both possible deployment scenarios on this link are analyzed. If not otherwise stated, an uncoded system with BPSK modulation is applied.

Using the relation for link budget, (1), and implementing corresponding path-loss models for AP – RS (see eq. (2)) and RS – ST (see eq. (3)) links, we have first obtained results on achievable ranges for both DL and UL communications in IEEE 802.11ah dual-hop DF relay systems, for scenarios without multipath fading. The graphs on Fig. 3 actually present the minimum among signal powers received at each of the two hops, given as a function of the RS - ST distance, for the chosen AP - RS distance of 500m. The Figure's legend has the following meanings: "macro-macro depl." denotes either DL or UL scenario when on both hops the path-loss model for the macro deployment is used; "macro-pico depl." denotes DL scenario where macro deployment is assumed for the first hop, while the pico deployment is on the second hop; "pico-macro depl." is introduced for UL communication, with the pico path-loss model on the first hop and the macro deployment model on the second hop.

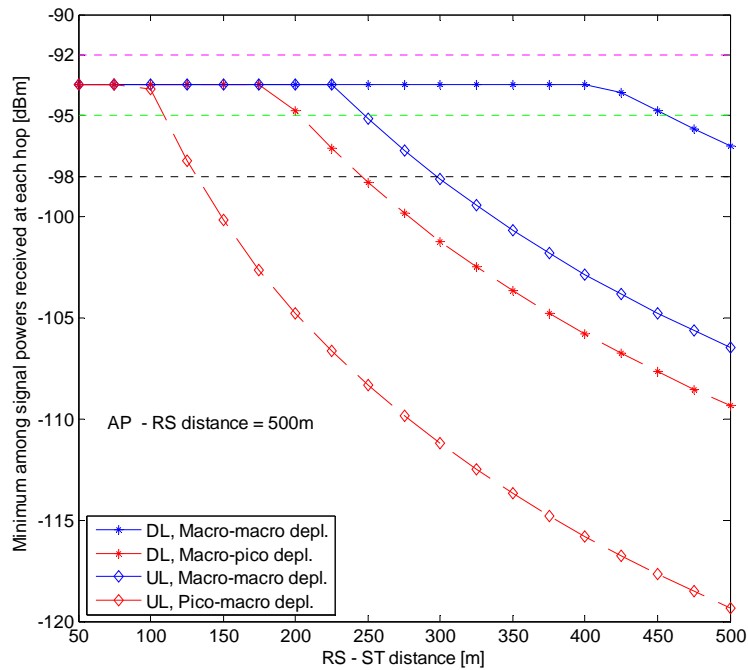

Fig. 3. Achievable ranges in IEEE 802.11ah relay system, taking into account path-loss only

The obtained achievable ranges for the direct AP – ST communication in the path-loss only scenario, previously presented in Fig. 1, are now extended for the chosen AP – RS distance of 500m. From Fig. 3, it can be seen that the target range between AP and ST of 1km is outreached on DL for the macro-macro deployment scenario and MCS10 implemented. The maximum achievable communication range on UL is about 840m, for the macro-macro deployment scenario and MCS10 scheme. For the chosen AP - RS distance, MCS0 can also be implemented in all four presented communication scenarios, which will result in 30-80m smaller range compared to the case with MCS10 implemented, depending on the deployment scenario on hops, and whether UL or DL is considered.

With the same methodology applied for analyzing the IEEE 802.11ah system without relays, the presence of multipath fading on both hops is added for further considerations. Multipath fading significantly influences the results on achievable ranges, all depending on the expected end-to-end outage probability, $P_{out\_tot}$. Using (4), and the corresponding *FM* terms for the assumed fading statistics on the hops, we have calculated the minimum among signal powers received at each hop, which for DL communication denotes lower of signal powers received at RS and ST, while for UL communications denotes lower of the signal powers received at RS and AP. Figure 4 gives results on DL and UL achievable ranges, for the expected end-to-end outage probability of $P_{out\_tot}$=0.1, and for the AP – RS distance of 400m. *FM* derivation for the assumed Rician fading channel is given in [16]. Using that, and the fact that the per-link outage ($P_{out}$) is equal to 5%, for the expected $P_{out\_tot}$ of 10%, we have obtained *FM* on the AP - RS link to be 4.5dB. On the RS - ST link, with Rayleigh fading statistics, *FM* is equal to 12.89dB (given in Table 5).

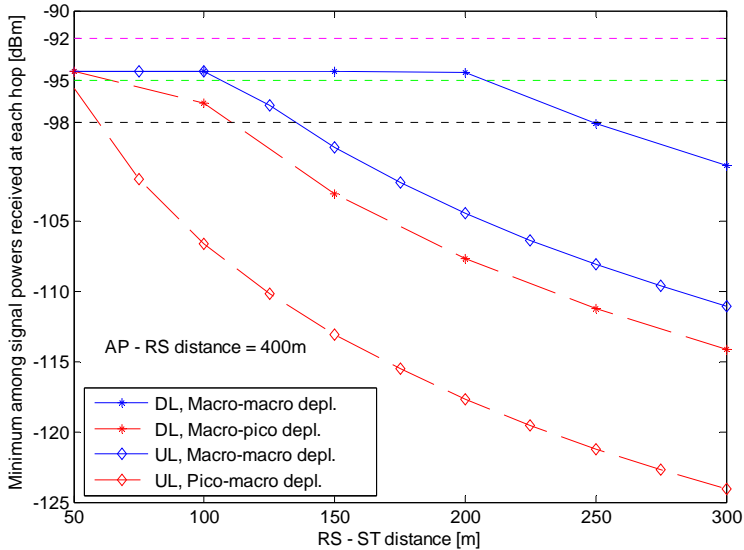

Fig. 4. Achievable ranges in IEEE 802.11ah relay system in multipath fading scenario

From Fig. 4, it can be concluded that the maximum achievable DL range on the RS - ST link is equal to 250m for MCS10, thus resulting in the maximum distance between AP and ST of 650m (400m + 250m), in the macro-macro deployment case. When RS is not placed 15m above rooftop, i.e. for the pico deployment scenario on the RS – ST link, maximum achievable DL range becomes 510m (400m+110m). Further on, the ST - RS uplink communication allows the maximum range of about 140m for the macro deployment, and 60m for the pico deployment, both for MCS10, which are extended by 400m distance between AP and RS.

It is evident that lowering the required level of the end-to-end link outage probability, a lower *FM* on both links in the dual-hop relay system would be obtained, which turns into greater achievable range. Fig. 5 presents DL communication range in the IEEE 802.11ah DF relay system with $P_{tx}$=10mW, in

___

cases where end-to-end link outages are $P_{out\_tot}$ =0.2 and $P_{out\_tot}$ =0.4. Table 5 gives information on $P_{out}$ per link and corresponding fade margins on each link in the chosen scenarios.

Table 5. *FM* values

| | **AP - RS link** | | **RS - ST link** | |
|---|---|---|---|---|
| $P_{out\_tot}$ | $P_{out}$ | *FM* [dB] | $P_{out}$ | *FM* [dB] |
| 20% | 10% | 3 | 10% | 9.77 |
| 40% | 20% | 2 | 20% | 6.51 |

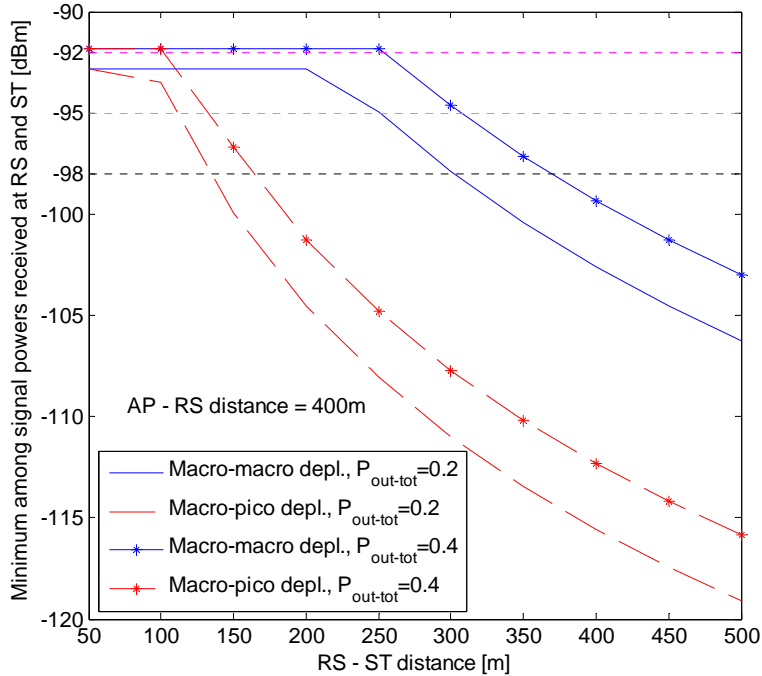

Fig. 5. Achievable DL ranges in IEEE 802.11ah relay system, in multipath fading scenario, for different $P_{out\text{-}tot}$ values

From Fig. 5, it can be seen that maximum achievable ranges on the RS - ST link are about 370m and 300m, for $P_{out\_tot}$=0.4 and $P_{out\_tot}$=0.2, respectively, both for MCS10 and the macro deployment scenario on this link. Moreover, it can be concluded that the AP - RS distance can be increased above the taken 400m, while still keeping the received signal power level above MDS at RS for MCS10. Of course, the ranges for the pico deployment scenario on the RS - ST link are far below the ones in the macro deployment scenario, and they are about 190m and 150m, for $P_{out\_tot}$=0.4 and $P_{out\_tot}$=0.2, respectively, for MCS10.

All presented analyses on achievable ranges in IEEE 802.11ah systems show positive effects of relay implementation, which directly contributes towards creating conditions for considering the new WLAN standard as an interesting solution for IoT and M2M based applications. To that aim, it is shown that the position of AP and RS should be carefully chosen, in order to meet the requirements for good communication link between them, i.e. to enable long range communications.

The benefits of RS implementation for range extension in IEEE 802.11ah systems are especially important for UL communication. Namely, for plethora of existing and foreseen IoT and M2M based applications, a complete traffic, or most of the traffic is transferred through the uplink. These end-devices (nodes, or stations) in most cases would be battery powered, thus requiring a low power transmission for extended battery life, and, at the same time, an achievement of the longest possible communication ranges. Such conditions are common for all world regions, and transmit power levels of 0dBm, like we assumed in our analysis, will be mostly used. Having this in mind, it is sure that IEEE 802.11ah will rely on implementation of relay stations, in order to prevail in performances over competing IEEE 802.15.4 solutions.

## 4 Data rate analysis

IEEE 802.11ah standard specifies the target data rate of at least $10^5$b/s, which should be attained for all stations receiving signals above the MDS level. It is clear that the most robust MCS schemes, achieving the greatest communication range, actually have the lowest data rate among all MCSs. Thus, in our further analyses of IEEE 802.11ah systems we put focus on the maximum communication ranges achieved, while the target data rate of at least $10^5$b/s is fulfilled. We consider the MCS0 scheme with $B$=1MHz, which is the second most robust MCS scheme in terms of achievable range, and it has twice higher data rate than MCS10. This is due to the fact that in terms of achievable data rate, this scheme is two times better than the MSC10 scheme, since the repetition coding in MSC10, while enabling the range extension, lowers twice the data rate.

We first analyze achievable data rates for the direct AP – RS communication (no relay), for the given packet error rate (PER), and in the presence of Rayleigh fading that may cause the link outage. The relation between data rate and PER can be derived using the expression for signal-to-noise ratio (SNR) at the receiver side:

$$SNR = \frac{S}{N} = \frac{S}{N_0 B} = \frac{E_b}{N_0 B T_b} = \frac{E_b R}{N_0 B} \qquad (9)$$

where $S$ represents the received signal power, which corresponds to the linear value of $P_{rx}$ from (4). $E_b$ is the bit energy, $T_b$ denotes the bit duration, $R$ is the data rate, $B$ is the receiver system bandwidth. The noise power, $N$, is given as:

$$N = N_0 B = k T_0 F B \qquad (10)$$

with $N_0$ denoting the noise spectral density, $k$ being Boltzmann's constant, $T_0$ is the receiver temperature expressed in Kelvins (K), and $F$ is the receiver noise figure.

If dB values are introduced in (9), it becomes:

$$\left(\frac{S}{N}\right)_{dB} = \left(\frac{E_b}{N_0}\right)_{dB} + 10\log_{10}\left(\frac{R}{B}\right) \qquad (11)$$

Combining relations (4) and (11), an expression for the maximum achievable system data rate (in dB) is obtained, being a function of distance $d$ (in m) between transmitter and receiver:

$$\left(R(d)\right)_{dB} = P_{tx} + G_{tx} - PL(d) - FM + G_{rx} - \left(E_b / N_0\right)_{dB} - N_0 \qquad (12)$$

Relation (12) gives an insight in maximum achievable data rate, for the required level of bit error rate (BER). Namely, the term $E_b/N_0$ is set to the value that can guarantee achievement of the required BER. For the BPSK modulation, which is implemented in MCS0 and MCS10, $E_b/N_0$, for the given BER, is derived from the expression:

$$BER = Q\left(\sqrt{\frac{2E_b}{N_0}}\right) \quad (13)$$

In WLAN systems, data are transferred in form of packets. If it is assumed that errors within the packets are independent and that they occur with an equal probability, the following relation between BER and the packet error rate (PER) applies:

$$PER = 1 - (1 - BER)^L, \quad (14)$$

where $L$ denotes the packet length in bits. In our analyses, for all considered packet lengths, it is taken that the end-to-end PER is equal to 10%.

Figure 6 presents the maximum achievable data rates for MCS0 scheme with $B=1$MHz, for the case of $L=4096$ bytes, PER=0.1 and $P_{out}=0.1$.

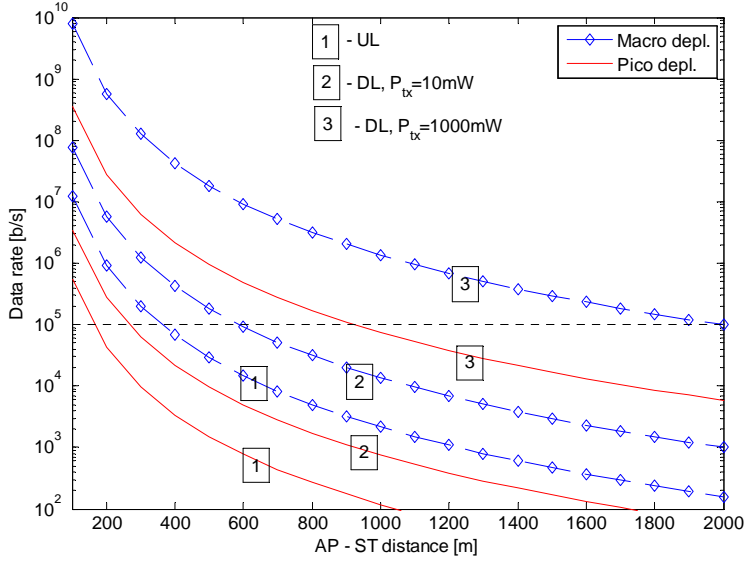

Fig. 6. Achievable data rate for MCS0 in IEEE 802.11ah system - Rayleigh fading scenario

From Fig. 6, it can be seen that, for the considered conditions, the target data rate of $10^5$b/s, at 1km distance between AP and ST is achieved on DL for the system having $P_{tx}=1$W in the case of macro deployment. For the pico deployment scenario, the system with $P_{tx}=1$W achieves DL range of more than 900m, with the data rate of $10^5$ b/s. For the system having $P_{tx}=10$mW, the target data rate of $10^5$b/s can be achieved for AP – ST distances lower than 600m and 270m for macro and pico deployment, respectively. When UL communication is considered, with ST having $P_{tx}=1$mW, the achievable ranges are about 180m and 380m, for macro and pico deployment, respectively, for the data rate of $10^5$ b/s.

Greater ranges, for the target data rate of $10^5$b/s, can be achieved if we assume a smaller packet size and/or implementation of coding, which would enable the same level of BER for a much lower $E_b/N_0$ ratio. However, as it was previously shown, the implementation of relay station is the most effective solution for range extension. Thus, we examine achievable data rates of dual-hop IEEE 802.11ah relay system for different packet sizes, and for both uncoded and coded systems.

For the considered relay transmission, based on the dual-hop DF relaying protocol, the end-to-end maximum achievable data rate can be obtained through:

$$R_{DF} = 0.5 \cdot \min\{R_1, R_2\} \tag{15}$$

where the term 0.5 is due to the transmission in two time-slots between the source of information and destination, and $R_1$ and $R_2$ denote the maximum achievable data rates on the first hop and second hop, respectively. Relation (15) actually shows that in the dual-hop DF relay system, the end-to-end data rate is upper limited by the worse of the two hops, i.e. with the one having the lower SNR.

From relations (12) - (14), it can be seen that the maximum achievable data rate depends on the PER value. In the considered IEEE 802.11ah dual-hop DF relay system erroneous packet is received at the destination if error occurs on the first hop ($PER_1$), or, if an error-free transmission has been accomplished on the first hop, but an error occurs on the second hop ($PER_2$). Thus, the expression for the end-to-end packet error rate in such a relay system can be written as:

$$PER_{DF} = PER_1 + PER_2(1 - PER_1) = PER_1 + PER_2 - PER_1 PER_2 \tag{16}$$

It is reasonable to assume that PER values on each hop will be significantly lower than 1, so we can neglect the term $PER_1 PER_2$ in the final $PER_{DF}$ expression, thus having an approximation:

$$PER_{DF} \approx PER_1 + PER_2 \tag{17}$$

In our further analysis, we have taken that PER values on both hops are the same, i.e. for the assumed $PER_{DF}$=0.1, $PER_1$=$PER_2$=0.05.

Figure 7 gives plots on the maximum achievable data rates on DL in the IEEE 802.11ah dual-hop DF relay system, for different packet lengths, $L$, when $P_{tx}$=10mW for both AP and RS. We have again assumed that the AP - RS channel has Rician fading statistics, and only the macro deployment scenario for this link is considered. It is taken that the AP – RS distance is 850m. Other significant parameters for the considered scenario are: MCS0 with $B$=1MHz and $P_{out-tot}$=0.1.

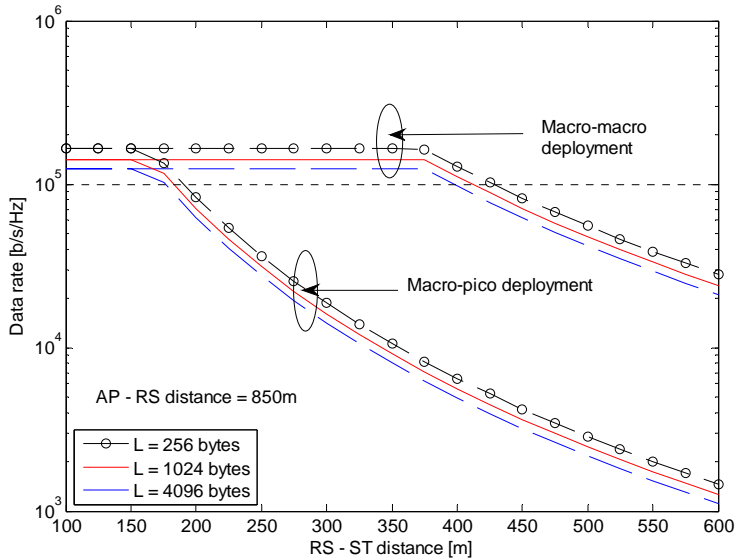

Fig. 7. Achievable DL data rate for MCS0 in IEEE 802.11ah relay system

The smaller packet size actually needs a lower $E_b/N_0$ ratio for achieving the required BER value, which then reflects to a higher data rate (see (12)). Small packets of $L$=256 bytes may be used in different IoT and M2M based applications.

From Fig. 7 it can be seen that, for the target data rate of $10^5$ b/s, the achievable range on RS - ST link is more than 420m in case of the macro deployment and $L$=256 bytes, and about 190m for the pico deployment. This means that even for the pico deployment on the RS - ST link, the goal of the communication range up to 1km with data rate of $10^5$b/s is achieved, as the chosen AP-RS distance is 850m.

If UL communication process is considered, Figure 8 shows that the uncoded MCS0 scheme for $L$=256 bytes, can achieve ranges of 120m and 270m on the first hop (ST - RS link), for pico and macro deployments on this link, respectively, all considering the target data rate of $10^5$ b/s. Having that the RS - AP distance is 850m (and can be a little bit more extended, while still achieving the $10^5$ b/s target data rate), then we can conclude that even in the uplink communication the desired data rate of $10^5$b/s can be achieved for the distances between ST and AP of up to 1km. This communication distance can be far more extended if coded MCS0 scheme is used. Thus, for example, a low density parity check (LDPC) decoder has approximately 8dB coding gain for BPSK modulation and 1/2 code rate, meaning that 8dB lower $E_b/N_0$ ratio for the same PER value per link is required [18]. This is the data we used in our analysis presented in Fig. 8. All the other relevant parameters are the same as in previous scenarios.

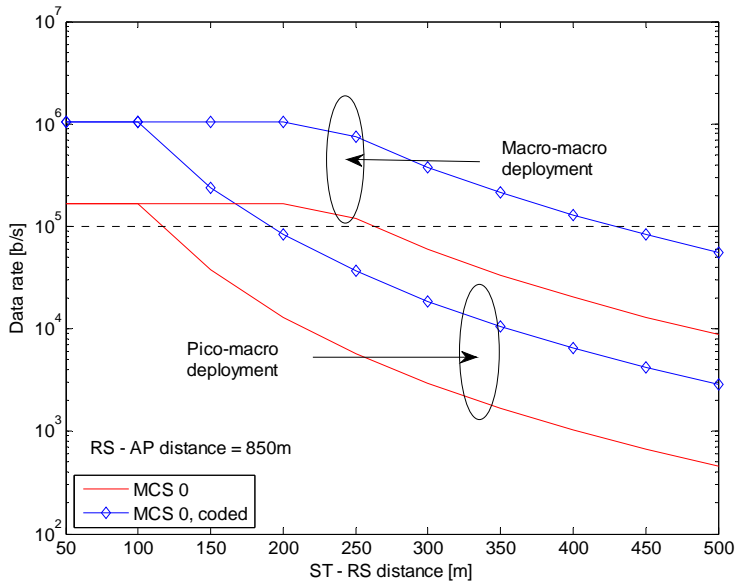

Fig. 8. Achievable UL data rate for MCS0 in IEEE 802.11ah relay system

The previously presented results on achievable data rates in IEEE 802.11ah systems for MCS0 with $B$=1MHz, show that, through the implementation of RS, with a carefully chosen position of AP, the target data rate of $10^5$ b/s can be achieved at AP – ST distances up to 1km, both on DL and UL. This means that whenever the signal is above the MDS level in IEEE 802.11ah relay system, then the communication process between AP and ST can be established through RS, with the data rate of $10^5$ b/s, or higher. It is shown that if any type of coding is used, or for the smaller end-to-end link outage, than these achievable ranges, for the target data rate of $10^5$b/s, can be further extended.

## 5 BER performance analysis

Having an insight in the level of range extension attainable through implementation of relay stations in IEEE 802.11ah systems, we wanted to examine and compare BER performances of the most robust MCS schemes, in scenarios where they will be used for establishing communications at the greatest feasible distances. Taking that RS employment provides additional 400m in the communication link between AP and ST, in the following section we present BER of the dual-hop IEEE 802.11ah relay system, obtained through Monte Carlo simulations. We analyze uncoded MCS0 and MCS10 schemes, meaning that we present BER performances of BPSK modulated dual-hop relay systems, with 1MHz channel bandwidth, which actually uses 24 data carriers separated by 31.25kHz for data transmission.

### 5.1 Simulation parameters

We model both DL and UL communication processes of the uncoded dual hop IEEE 802.11ah DF relay system, with the focus on parameters' values adjusted to European regulations. This assumes that the AP transmit power is set to $P_{tx}$=10mW (10dBm), while for the 802.11ah station 1mW (0dBm) transmit power is defined.

Rician fading channel with $K$ factor of 9dB is modeled on AP – RS link, while the path-loss on this link is calculated assuming the macro deployment scenario, for both DL and UL communications. The Rayleigh fading on RS - ST link is assumed, and both possible path loss models on this link are analyzed. We used Typical Urban model for multipath fading on both links, [15]. For RS with the fixed position and stationary end-station (ST), Doppler shift is equal to 0Hz for each scenario considered. Additive white Gaussian Noise (AWGN) is assumed at each receiving communication device, with the noise power of -145.22dB. Noise figures of AP and RS are equal to 3dB, while for ST it is equal to 5dB. Antenna gains at AP and RS are 3dBi, and ST has 0dBi antenna gain.

In these analyses we have assumed that RS and AP on UL, or RS and ST on DL, have the perfect corresponding channel knowledge, so channel estimations are not included in the simulation model.

Repetition coding is implemented in time domain for MCS10, assuming that the same OFDM symbol is sent in two consecutive symbol intervals, and then two received symbols are combined in the receiver, after the channel equalization.

### 5.2  Results

BER results for the DL communication, presented as a function of RS – ST distance in dual-hop IEEE 802.11ah relay system are presented in Fig. 9, while BER results for the UL communication are given in Fig. 10.

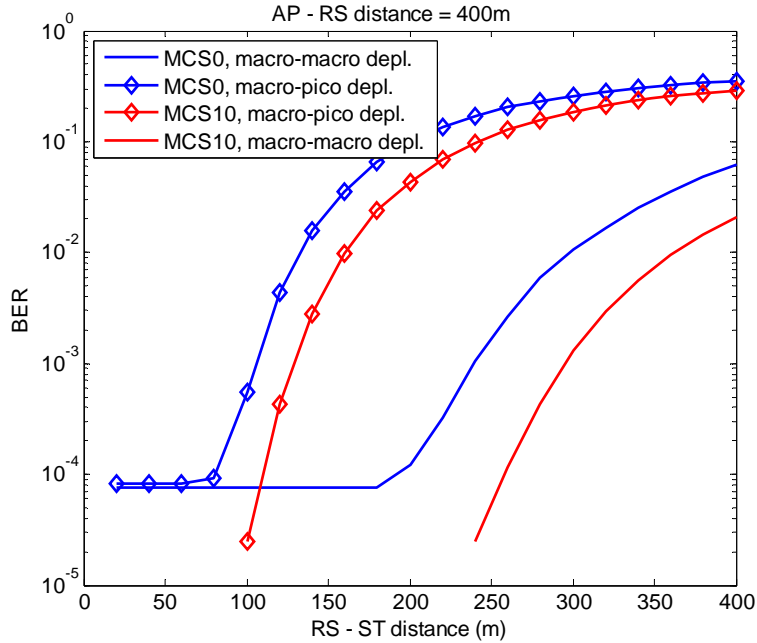

Fig. 9. DL BER performances of IEEE 802.11ah relay system

From the results presented in Fig. 4, we have seen that, for the same assumed multipath scenario as taken in this BER analyses, and for $P_{out-tot}$=0.1, the achievable ranges on DL for the IEEE 802.11ah relay system in the macro-macro deployment scenario are about 650m (400m AP – RS distance + 250m RS – ST distance) and 610m (400m+210m), for MCS10 and MCS0, respectively. From Fig. 9, we can now read the expected BER values for these greatest achievable ranges in the given conditions, thus having BER of $5 \cdot 10^{-5}$ for MCS10 at 250m RS – ST distance, and about $1,5 \cdot 10^{-4}$ for MCS0 at 210m distance between RS and ST.

For the macro-pico deployment scenario, the achievable DL ranges for the same assumed other conditions are significantly lower, and are equal about 510m (400 + 110m) and 465m (400m + 65m), for MCS10 and MCS0, respectively. Corresponding BER values at the edge of the RS coverage zone in the assumed scenario are equal to $10^{-4}$ for MCS10 and $9 \cdot 10^{-5}$ for MCS0.

When the UL communication is considered, we have seen that the maximum achievable ranges for the macro-macro deployment scenario in IEEE 802.11ah relay system, and for the assumed conditions, are equal to 540m (140m ST – RS distance + 400m RS – AP distance) and 510m (110m + 400m), for MCS10 and MCS0, respectively (see Fig. 4). Corresponding BER values obtained at these distances are about $10^{-5}$ for MCS10 at 540m, and $7 \cdot 10^{-5}$ for MCS0 at distance of 510m (Fig. 10).

In the pico-macro UL communication scenario, for the maximum achievable range of 460m, the BER value close to $10^{-5}$ can be expected. The second largest UL achievable range of 440m in this scenario, assuming employment of the MCS0 scheme, brings the BER value close to $10^{-4}$.

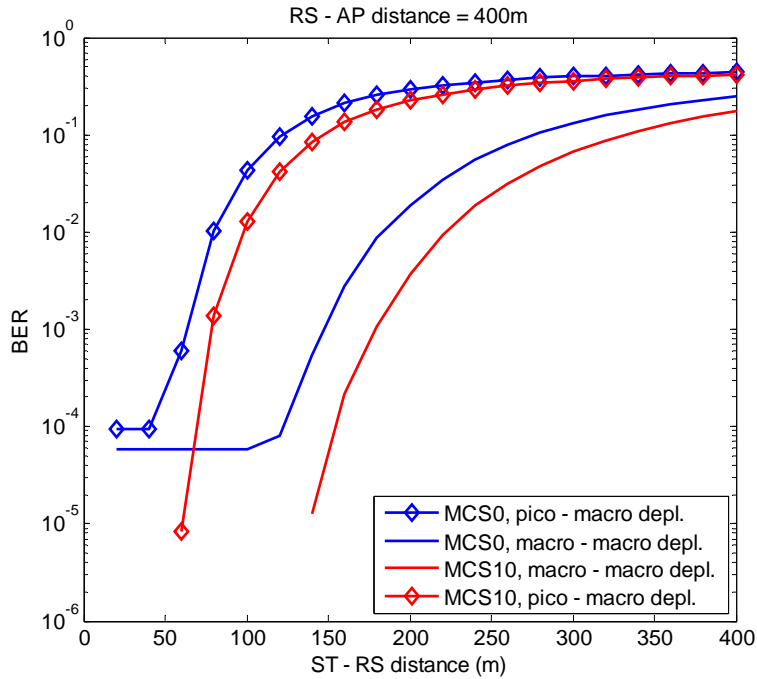

Fig. 10. UL BER performances of IEEE 802.11ah relay system

Both Fig. 9 and Fig. 10 show the benefits that the repetition coding (MCS10) brings to IEEE 802.11ah relay systems in terms of the BER performance improvement on DL and UL. Thus for example, in the case of DL communication and macro-macro deployment, at the RS – ST distance of 250m, MCS10 attains BER value of $5·10^{-5}$, while the MCS0 scheme has BER value equal to $1.7·10^{-3}$. In UL communication process, for the same deployment scenario, and at the same ST – RS distance of 250m, the MCS10 scheme achieves the BER value of about $2.2·10^{-2}$, while the MCS10 scheme has BER equal to $6.5·10^{-2}$.

At the end, the significance of carefully choosing the position of both AP and RS for the overall performances of IEEE 802.11ah systems is also confirmed through analysis of BER performances. Namely, BER results in the macro-macro scenario are much better than the corresponding ones in the macro-pico (or pico-macro on UL) deployment scenario. Thus, for example, we have that, in the DL communication and for the MCS10 scheme, BER value of $10^{-4}$ in the macro-macro deployment scenario is outperformed at RS – ST distances shorter than 260m, while for the macro – pico deployment, this BER value is outperformed only at distances shorter than 110m.

## 6 Conclusions

Telecommunication market is undergoing through major changes, meaning that the telecommunication industry and service providers must seek for efficient and cost-effective solutions for enabling implementation of IoT and M2M based applications, in order to remain competitive and to meet customers' demands. There is no doubt that this is the most promising segment of telecommunication market, with the explosive growth, which is reflected through tens of billions of different devices to be interconnected in the next couple of years. Being the leader in defining WLAN solutions, the IEEE 802.11 standardization group has defined the "ah" amendment as the extended range WLAN, operating at bands below 1GHz, and offering lower data rates, narrow bandwidth communications for IoT and M2M based applications. Besides using the channel at sub-1GHz frequency band, which has much better propagation characteristics than channels at 2.4GHz and 5GHz used by most of the existing IEEE 802.11 standards, the IEEE 802.11ah standard incorporates different physical and link layer solutions for achieving long-range communications and serving up to 6000 nodes by a single access point. The defined goal is to provide communication at distances up to 1km, with at least 100 kb/s data rate.

Having that the maximum allowed transmit power for the access point (AP) in some regions is limited to 10mW, we have shown that in this case the targeted range cannot be achieved in real case scenarios, assuming the presence of multipath fading, even when the most robust modulation and coding schemes (MCSs) are used. Thus, in this paper we provide for the first time comprehensive analysis on the level of a range extension achievable through the implementation of relay stations (RSs), which is shown to be the most effective solution for attaining long range communications in IEEE 802.11ah systems. Moreover, we examined whether the considered dual-hop relay systems achieving the longest range through the implementation of the most robust MCSs, can also meet the required data rate of 100kb/s.

In the presented analyses of IEEE 802.11ah two-hop DF relay systems, we have assumed real case scenarios, where positions of AP and RS are chosen in such a way to ensure a link with line of sight communication between them (i.e. Rician fading channel), while for the RS - ST channel we assumed Rayleigh fading statistics. All the conducted analyses and presented results have shown that, even with 10mW of the AP (and RS) transmit power, IEEE 802.11ah systems may be a very interesting and competitive solution for IoT and M2M based applications, if relay stations are implemented. By carefully choosing the position of AP and RS, the expected range on DL will be up to 800m, while on UL it will be about 550m, all for the end-to-end link outage probability of 0.1. On the other side, the targeted data rate of 100kb/s will be outreached even in the UL communication process, whenever the signal level at the destination is equal to, or above MDS.

Operating in the license-exempt frequency band, with a flexibility in providing different data rates, an ability to provide connection of up to 6000 devices on a single AP, and with communication ranges which are much higher than the ones achievable in systems based on the IEEE 802.15.4 standard, IEEE 802.11ah will certainly find its place on the future telecommunication market of IoT and M2M based applications.

**Acknowledgement**

This work has been supported by the Ministry of Science of Montenegro and the HERIC project through the BIO-ICT Centre of Excellence (Contract No. 01-1001).

**References**

1. Evans, D., (2011). The Internet of Things – how the next evolution of the Internet is changing everything, *Cisco Internet Business Solutions Group (IBSG)*.
2. Park, M., (2015). IEEE 802.11ah: Sub-1-GHz License-Exempt Operation for the Internet of Things. *IEEE Communications Magazine*, vol. 53, no. 9, pp. 145-151, September 2015
3. IEEE Standard 802.11ah, 2016 (2017). 802.11ah-2016 - IEEE Approved Draft Standard for Information Technology-Telecommunications and Information Exchange Between Systems-Local and Metropolitan Area Networks-Specific Requirements-Part 11: Wireless LAN Medium Access Control (MAC) and Physical Layer (PHY) Specifications: Amendment 2: Sub 1 GHz License Exempt Operation, February 2017.
4. Aust, S., Ito, T., (2012). Sub 1GHz wireless LAN propagation path loss models for urban smart grid applications. *International Conference on Computing, Networking and Communications (ICNC)*, pp. 116 – 120, Jan. 30 2012-Feb. 2 2012.
5. Hazmi, A., Rinne, J., and Valkama, M., (2012). Feasibility Study of IEEE 802.11ah Radio Technology for IoT and M2M use Cases. *GC'12 Workshop: Second International Workshop on Machine-to- Machine Communications 'Key' to the Future Internet of Things*, pp. 1687 – 1692.
6. Adame, T., Bel, A., Bellalta, B., Barcelo, J., Oliver, M., (2014). IEEE 802.11ah: the WiFi approach for M2M communications. *IEEE Wireless Communications*, Volume: 21, Issue: 6, pp. 144 – 152.
7. Khorov, E., Lyakhov, A., Krotov, A., Guschin, A., (2015). A survey on IEEE 802.11ah: An enabling networking technology for smart cities. *Computer Communications,* Volume 58, pp. 53–69.
8. Aust, S., Prasad, R., Niemegeers, I. G. M. M. (2015). Outdoor Long-Range WLANs: A Lesson for IEEE 802.11ah. *IEEE Communications Surveys & Tutorials,* Volume: 17, Issue: 3, pp. 1761 - 1775.
9. Sun, W., Choi, M., and Choi, S., (2013). IEEE 802.11ah: A Long Range 802.11 WLAN at Sub 1 GHz. *River Publisher Journal*, Article 2245 800X 115
10. Domazetović, B., Kočan, E., Mihovska, A., (2016). Performance evaluation of IEEE 802.11ah systems. *In Telecommunications Forum (TELFOR),* 24$^{th}$, 2016., pp. 1-4.
11. Argyriou, A., (2015). Power-Efficient Estimation in IEEE 802.11ah Wireless Sensor Networks with a Cooperative Relay. *IEEE International Conference on Communications (ICC),* 2015., pp. 6755 – 6760.
12. Venkatasubramanian, S. N., Haneda, K., Yamamoto, K., (2015). System-level Performance of In-Band Full-Duplex Relaying on M2M Systems at 920 MHz. *IEEE Vehicular Technology Conference (VTC Spring)*, 81$^{st}$, 2015.  pp. 1-5
13. Park, M., (2013). IEEE 802.11 Wireless LANs Specification Framework for TGah doc. *IEEE 802.11-11/1137r15doc.May 2013*.
14. Aust, S., Prasad, R., Niemegeers I.G.M.M., (2012). IEEE 802.11ah: Advantages in standards and further challenges for sub 1 GHz Wi-Fi. *IEEE International Conference on Communications (ICC)*, 2012, pp. 6885–6889
15. Ponnampalam, V., Wang, V., Porat, R., (2011). IEEE P802.11 Wireless LANs TGah - Outdoor Channel Models. *IEEE 802.11-11/0760r2, 2011.*
16. Molisch, A. F., (2011). Wireless Communications, 2nd edition, Wiley, United Kingdom
17. IEEE 802.11ah/D5.0, "Draft for Information Technology — Telecommunications and Information Exchange between Systems — Local and Metropolitan Area Networks — Specific Requirements — Part 11: Wireless LAN Medium Access Control (MAC) and Physical Layer (PHY) Specifications — Amendment 6: Sub 1 GHz License Exempt Operation."
18. Richardson, T. and Urbanke, R., (2011). Efficient encoding of lowdensity parity-check codes, *IEEE Trans. Inform. Theory*, vol. 47, no. 2, pp. 638- 656, Feb. 2001.